\documentclass{article}
\usepackage{xcolor}
\usepackage{multirow}
\usepackage{amsfonts} 
\usepackage{spconf,amsmath,graphicx}
\usepackage{float}
\usepackage{hyperref}

\title{Emphasized Non-Target Speaker Knowledge in Knowledge Distillation \\for Automatic Speaker Verification}

\name{Duc-Tuan Truong$^1$, Ruijie Tao$^{2, \dag}$\thanks{$\dag$ Corresponding author.}, Jia Qi Yip$^{1,3}$, Kong Aik Lee$^{4}$, Eng Siong Chng$^1$}
\address{$^1$Nanyang Technological University, Singapore\quad$^2$National University of Singapore, Singapore \\ $^3$Alibaba Group\quad$^4$The Hong Kong Polytechnic University, Hong Kong \\}
\usepackage{fancyhdr}

\fancypagestyle{firstpage}{
    \fancyhf{}
    \fancyfoot[L]{\footnotesize\textcopyright 2024 IEEE. Personal use of this material is permitted. Permission from IEEE must be obtained for all other uses, in any current or future media, including reprinting/republishing this material for advertising or promotional purposes, creating new collective works, for resale or redistribution to servers or lists, or reuse of any copyrighted component of this work in other works.}
 }

\begin{document}
\thispagestyle{firstpage}
%
\maketitle

\begin{abstract}
Knowledge distillation (KD) is used to enhance automatic speaker verification performance by ensuring consistency between large teacher networks and lightweight student networks at the embedding level or label level. However, the conventional label-level KD overlooks the significant knowledge from non-target speakers, particularly their classification probabilities, which can be crucial for automatic speaker verification. In this paper, we first demonstrate that leveraging a larger number of training non-target speakers improves the performance of automatic speaker verification models. Inspired by this finding about the importance of non-target speakers' knowledge, we modified the conventional label-level KD by disentangling and emphasizing the classification probabilities of non-target speakers during knowledge distillation. The proposed method is applied to three different student model architectures and achieves an average of 13.67\% improvement in EER on the VoxCeleb dataset compared to embedding-level and conventional label-level KD methods\footnote{Code and models are available at \href{https://github.com/ductuantruong/enskd}{github.com/ductuantruong/enskd}}.

\end{abstract}
\section{Introduction}
\label{sec:intro}

Automatic speaker verification (ASV) is the process of authenticating an individual's claimed identity based on voice characteristics. By leveraging large-scale neural networks trained on abundant unlabelled speech data, self-supervised learning (SSL) models have revolutionized various speech processing tasks \cite{ssl_app_asr,ssl_app_kw,ssl_app_age_2}, including ASV \cite{ssl_app_sv, ssl_app_sv2}. However, these models are computationally expensive. To better utilize SSL models, knowledge distillation can be employed to transfer the robust speech representation to smaller student models. In ASV, KD encompasses two common approaches: one is embedding-level method \cite{emb_kl_kd_1,emb_kd_2,emb_kd_4,emb_kd_5}, which attempts to make student models mimic the teacher’s intermediate feature embedding by reducing the distance between representation spaces; the other is label-level method \cite{emb_kl_kd_1, kl_kd_2}, which focuses on minimizing the Kullback–Leibler divergence between the output probabilities of the teacher and student networks.

In the training step of an ASV model, the objective is to classify input speech into target speaker (the ground-truth speaker) and avoid assigning it to non-target speakers (incorrect speakers). While the importance of the target speaker is evident, non-target speakers can also enhance the model's discriminability since there would be numerous non-target speakers sharing similar voice characteristics with the target speaker. In past studies, \cite{nontarget_speaker} compared ASV models performance trained on two training sets with different numbers of speakers but the same number of utterances, and found that a larger number of speakers improved the performance. Similarly, in face recognition, \cite{nontarget_face} also observed that an increasing number of training non-target classes improved model performance within a fixed-size training set. Building on these observations, we hypothesize that integrating knowledge from non-target speakers can enhance ASV model performance. However, the conventional label-level KD considers correlations among the teacher's output probabilities of all speakers, the importance of non-target speakers' probabilities can be overshadowed by the target speaker with high classification confidence in the teacher model. Based on this hypothesis, the conventional label-level KD approach for ASV can be improved by emphasizing the knowledge of non-target speakers. 

To validate the assumptions above, this paper initially shows an experiment illustrating the importance of non-target speakers in ASV. When the number of training utterances remains the same, we observe that an increasing number of non-target training speakers leads to better results. Based on this observation, we investigate the significance of non-target speakers in the conventional label-level KD for ASV models. Following Decoupled Knowledge Distillation (DKD) \cite{dkd}, we segregate the output classification probabilities of the teacher and student models into two distinct probabilities of target and non-target speakers. Subsequently, the probabilities of non-target speakers are emphasized during KD using a specific weight. We utilize the large-scale SSL model WavLM-TDNN \cite{wavlm} as our teacher model and employ three different network architectures: x-vector \cite{xvector}, ResNet34 \cite{resnet34}, and CAM++ \cite{campplus} as student models. Our experiments show that DKD with an emphasis on the non-target speakers' output probabilities, outperforms both embedding-level and conventional label-level KD methods across student models. 


\section{Methodology}
\label{sec:method}
\subsection{The impact of non-target speakers for ASV}
\label{sec:study}
\begin{figure}[t]
  \centering
  \includegraphics[width=0.42\textwidth]{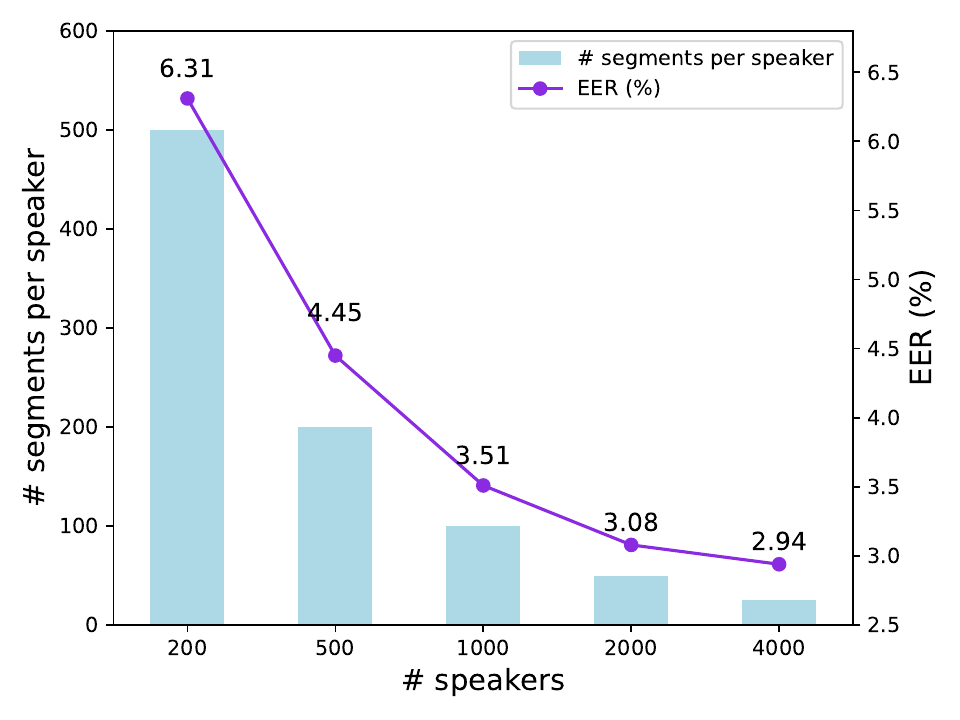}
  \caption{Vox1-O results (EER \%) of x-vector model trained on a fixed number of utterances but varying numbers of speakers.} 
  \label{study}
\end{figure}

To validate the hypothesis that a larger set of non-target speakers benefits ASV models, we conducted a toy experiment. We trained the x-vector model using a fixed 100,000 training utterances of the VoxCeleb 2 dev set \cite{vox}. These training utterances are evenly distributed among each training speaker, hence increasing the number of speakers will lead to fewer training utterances per speaker. As depicted in Figure \ref{study}, the performance of the x-vector model consistently improves with an increasing number of speakers in the training set. This indicates that involving more non-target speakers enhances the model's ability to distinguish the target speaker from others. Inspired by this finding, we further extract and emphasize non-target speaker knowledge during the knowledge distillation process. 

\begin{figure*}[h]
  \centering
  \includegraphics[width=0.83\textwidth]{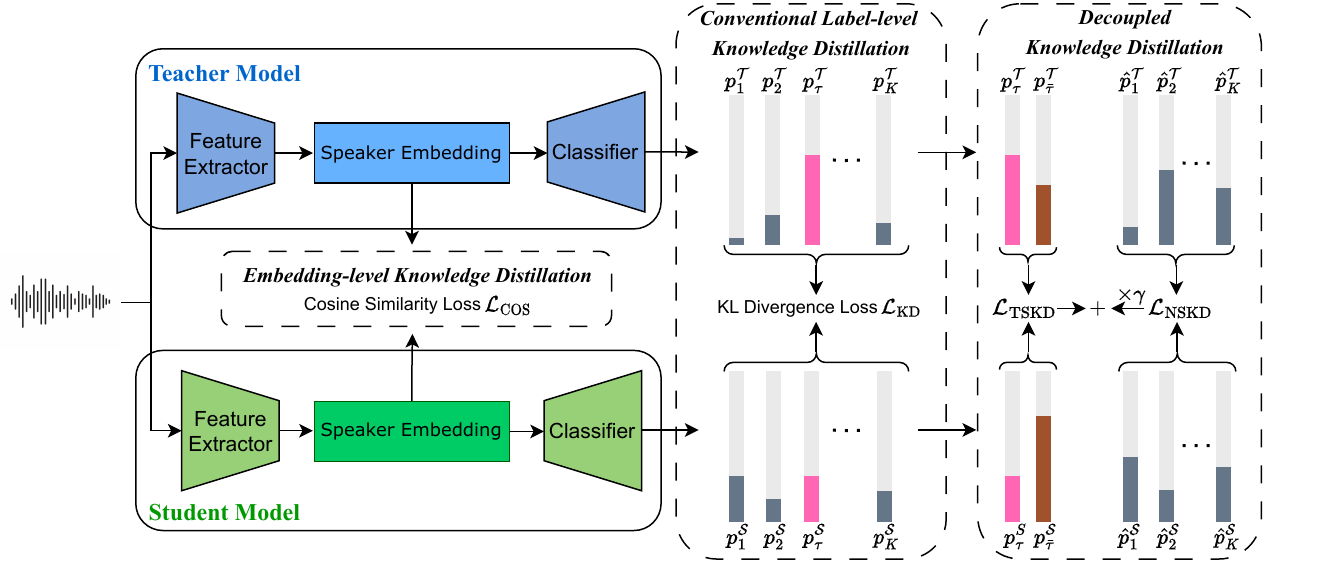}
  \caption{Our Decoupled Knowledge Distillation (DKD) with an emphasis on non-target speaker knowledge in comparison with the embedding-level knowledge distillation (using cosine distance loss $\mathcal{L}_{\text{COS}}$) and the conventional label-level knowledge distillation (using Kullback–Leibler divergence loss $\mathcal{L}_{\text{KD}}$). $\mathcal{T}$, $\mathcal{S}$, $K$, and $\tau$ denote the teacher model, the student model, the number of training speakers, and the target speaker, respectively. $p_{i}$, $p_{\bar{\tau}}$, and $\hat{p}_i$ are respectively defined as Eq.(\ref{eq:softmax}) and Eq.(\ref{eq:nc_softmax}). $\mathcal{L}_{\text{TSKD}}$, $\mathcal{L}_{\text{NSKD}}$ and $\gamma$ are defined as Eq.(\ref{eq:bf_final_kld}) and Eq.(\ref{eq:final_dkd}), respectively.}
  \label{fig:main}
\end{figure*}

\subsection{Rethinking conventional label-level KD}
Following the reformulation of the conventional label-level KD for the computer vision task in \cite{dkd}, we interpret the conventional label-level KD loss for automatic speaker verification. In the training phase of the ASV model, the model's output is classification probabilities $\mathbf{p}$ over the set $\mathcal{K}$ of \textit{K} training speakers, in which the probability $p_i$ of the \textit{i}-th speaker is computed using the softmax function to transform the logits vector $z_i$ into a probability distribution as follows:
\begin{equation} \label{eq:softmax}
    p_i = \frac{e^{z_i}}{\sum_{j=1}^Ke^{z_j}} 
\end{equation}
In the conventional label-level KD, the student model tries to mimic the teacher model by minimizing the Kullback-Leibler Divergence $D_{\text{KL}}$ between the student ($\mathcal{S}$) and teacher ($\mathcal{T}$) output probability distributions. The $D_{\text{KL}}$ loss is defined as:
\begin{equation} \label{eq:kld}
    \mathcal{L}_{\text{KD}} = D_{\text{KL}}(\mathbf{p}^{\mathcal{T}}\Vert \mathbf{p}^{\mathcal{S}}) = \sum_{i\in K}p^{\mathcal{T}}_{i}\log(\frac{p^{\mathcal{T}}_{i}}{p^{\mathcal{S}}_{i}})
\end{equation}
where $\mathbf{p}^{\mathcal{T}}, \mathbf{p}^{\mathcal{S}}$ denote the output probabilities of the teacher and student networks, respectively. We further split the set $\mathcal{K}$ of indexes $\{i=1\dots K\}$ in $\mathcal{L}_{\text{KD}}$ into the target speaker $\tau$ and a set of non-target speakers $\mathcal{K}\backslash\{\tau\}$ as: 
\begin{equation} \label{eq:split_kld}   
    \mathcal{L}_{\text{KD}} = p_{\tau}^{\mathcal{T}} \log(\frac{p_{\tau}^{\mathcal{T}}}{p_{\tau}^{\mathcal{S}}})+\sum_{i\in \mathcal{K}\backslash\{\tau\}} p_i^{\mathcal{T}} \log(\frac{p_i^{\mathcal{T}}}{p_i^{\mathcal{S}}}) 
\end{equation}
We define the probability of classifying a speaker belonging to $\mathcal{K}\backslash\{\tau\}$ as $p_{\bar{\tau}}$, and the probability of predicting a specific non-target speaker $i\neq \tau$ over all non-target speakers as $\hat{p}_{i}$:
\begin{equation} \label{eq:nc_softmax}
    p_{\bar{\tau}} = \frac{\sum_{i\in \mathcal{K}\backslash\{\tau\}}e^{z_i}}{\sum_{j\in K}e^{z_j}},\quad \hat{p}_{i} = \frac{e^{z_{i}}}{\sum_{j\in \mathcal{K}\backslash\{\tau\}}e^{z_j}}
\end{equation}
From ({\ref{eq:softmax}}) and ({\ref{eq:nc_softmax}}), we replace $p_{i} = p_{\bar{\tau}}\hat{p}_{i}$ in ({\ref{eq:split_kld}}):
\begin{align} \label{eq:mid_kld}   
    \mathcal{L}_{\text{KD}} &= p_{\tau}^{\mathcal{T}} \log(\frac{p_{\tau}^{\mathcal{T}}}{p_{\tau}^{\mathcal{S}}})+p_{\bar{\tau}}^{\mathcal{T}}\sum_{i\in \mathcal{K}\backslash\{\tau\}} \hat{p}_i^{\mathcal{T}} \log(\frac{p_{\bar{\tau}}^{\mathcal{T}}\hat{p}_i^{\mathcal{T}}}{p_{\bar{\tau}}^{\mathcal{S}}\hat{p}_{i}^{\mathcal{S}}})\nonumber\\
    &= p_{\tau}^{\mathcal{T}} \log(\frac{p_{\tau}^{\mathcal{T}}}{p_{\tau}^{\mathcal{S}}}) + p_{\bar{\tau}}^{\mathcal{T}}\sum_{i\in \mathcal{K}\backslash\{\tau\}}\hat{p}_i^{\mathcal{T}}\log(\frac{p_{\bar{\tau}}^{\mathcal{T}}}{p_{\bar{\tau}}^{\mathcal{S}}}) \nonumber\\ & + p_{\bar{\tau}}^{\mathcal{T}}\sum_{i\in \mathcal{K}\backslash\{\tau\}} \hat{p}_i^{\mathcal{T}} \log(\frac{\hat{p}_i^{\mathcal{T}}}{\hat{p}_{i}^{\mathcal{S}}})
\end{align}
Since $p_{\bar{\tau}}^{\mathcal{T}}$, $p_{\bar{\tau}}^{\mathcal{S}}$ are independent to the class index \textit{i} and $\sum_{i\in \mathcal{K}\backslash\{\tau\}} \hat{p}_i^{\mathcal{T}} = 1$, we can simplify (\ref{eq:mid_kld}) to: 
\begin{align} \label{eq:bf_final_kld}
    \mathcal{L}_{\text{KD}} &= \underbrace{p_{\tau}^{\mathcal{T}} \log(\frac{p_{\tau}^{\mathcal{T}}}{p_{\tau}^{\mathcal{S}}}) + p_{\bar{\tau}}^{\mathcal{T}}\log(\frac{p_{\bar{\tau}}^{\mathcal{T}}}{p_{\bar{\tau}}^{\mathcal{S}}})}_{D_\text{KL}(\mathbf{b}^{\mathcal{T}}\Vert \mathbf{b}^{\mathcal{S}})} + p_{\bar{\tau}}^{\mathcal{T}}\underbrace{\sum_{i\in \mathcal{K}\backslash\{\tau\}} \hat{p}_i^{\mathcal{T}} \log(\frac{\hat{p}_i^{\mathcal{T}}}{\hat{p}_{i}^{\mathcal{S}}})}_{D_\text{KL}(\mathbf{\hat{p}}^{\mathcal{T}}\Vert \mathbf{\hat{p}}^{\mathcal{S}})}
\end{align}
From (\ref{eq:bf_final_kld}), the conventional label-level KD can be re-formulated into the sum of two terms: 1) Target Speaker Knowledge Distillation (TSKD) loss $\mathcal{L}_{\text{TSKD}}$: the $D_\text{KL}$ over the binary classification probability $\mathbf{b} \in \mathbb{R}^2$ of the target speaker and all non-target speakers, and 2) Non-Target Speaker Knowledge Distillation (NSKD) loss $\mathcal{L}_{\text{NSKD}}$: the $D_\text{KL}$ of the multi-class classification probability $\mathbf{\hat{p}} \in \mathbb{R}^{K-1}$ between $K - 1$ non-target speakers as shown in (\ref{eq:final_kld}) and Fig.\ref{fig:main}.
\begin{align} \label{eq:final_kld}
    D_\text{KL}(\mathbf{p}^{\mathcal{T}}\Vert \mathbf{p}^{\mathcal{S}}) = \underbrace{D_\text{KL}(\mathbf{b}^{\mathcal{T}}\Vert \mathbf{b}^{\mathcal{S}})}_{\mathcal{L}_{\text{TSKD}}} + (1 - p_{\tau}^{\mathcal{T}})\underbrace{D_\text{KL}(\mathbf{\hat{p}}^{\mathcal{T}}\Vert \mathbf{\hat{p}}^{\mathcal{S}})}_{\mathcal{L}_{\text{NSKD}}}
\end{align}
From the above equation, when the teacher model predicts the target speaker accurately, a large value of $p_{\tau}^{\mathcal{T}}$ results in a smaller $(1 - p_{\tau}^{\mathcal{T}})$, which leads to the suppression of $\mathcal{L}_{\text{NSKD}}$. This could potentially hinder the distillation of knowledge from non-target speakers in the label-level KD method.

\subsection{Decoupled Knowledge Distillation with an emphasis on non-target speaker knowledge}
In Section \ref{sec:study}, it was demonstrated that leveraging more non-target speaker knowledge can enhance the performance of ASV models. In other words, $\mathcal{L}_\text{NSKD}$ may play a crucial role in the knowledge transfer from the teacher to student models. Decoupled Knowledge Distillation (DKD) \cite{dkd} proposed a modification to remove the dependency factor $(1 - p_{\tau}^{\mathcal{T}})$ in (\ref{eq:final_kld}) by introducing hyperparameters to balance the $\mathcal{L}_\text{TSKD}$ and $\mathcal{L}_\text{NSKD}$. However, to place a greater emphasis on $\mathcal{L}_\text{NSKD}$, we adjusted the original DKD method by simply replacing $(1 - p_{\tau}^{\mathcal{T}})$ with the hyperparameter $\gamma$ in the following manner:
\begin{equation} \label{eq:final_dkd}
    \mathcal{L}_\text{DKD} = \mathcal{L}_{\text{TSKD}} + \gamma \mathcal{L}_{\text{NSKD}}
\end{equation}
Finally, the DKD loss $\mathcal{L}_\text{DKD}$ is combined with the classification loss to optimize the student model. Fig.~\ref{fig:main} illustrates the comparison between DKD with an emphasis on non-target speaker knowledge, embedding-level, and conventional label-level knowledge distillation. 

\begin{table*}[t]
\caption{Results on the VoxCeleb1 test sets. $COS$ and $KLD$ denote embedding-level and conventional label-level KD} 
\label{tab:main_result}
\centering
\begin{tabular}{ccccccc}
\hline
\multirow{2}{*}{\textbf{System}}                                                                & \multirow{2}{*}{\textbf{\textbf{\textbf{\begin{tabular}[c]{@{}c@{}}Params\\ (M)\end{tabular}}}}} & \multirow{2}{*}{\textbf{\textbf{\begin{tabular}[c]{@{}c@{}}FLOPs\\ (G)\end{tabular}}}} & \multirow{2}{*}{\textbf{\begin{tabular}[c]{@{}c@{}}Distillation\\ Method\end{tabular}}} & \multicolumn{3}{c}{\textbf{EER (\%) / minDCF}}   \\
                                                                                                &                                  &                               &                                                                                         & \textbf{Vox1-O} & \textbf{Vox1-E} & \textbf{Vox1-H} \\ \hline
\begin{tabular}[c]{@{}c@{}}\textit{Teacher model}\\ WavLM-TDNN \cite{wavlm}\end{tabular}                              & 316.62                          & $\sim$26                           & -                                                                                       & 0.383 /    \text{\quad-\quad}            & 0.480 /     \text{\quad-\quad}           & 0.986 /     \text{\quad-\quad}           \\ \hline
\multirow{4}{*}{\begin{tabular}[c]{@{}c@{}}\textit{TDNN-based} \\ \textit{Student model}\\ x-vector \cite{xvector}\end{tabular}} & \multirow{4}{*}{4.61}           & \multirow{4}{*}{0.53}        & -                                                                                       & 1.835 /       \text{\quad-\quad}      & 1.822 /     \text{\quad-\quad}   & 3.110 /     \text{\quad-\quad}   \\
                                                                                                &                                  &                               & $COS$                                                                & 1.760 / 0.189           & 1.742 / 0.185           & 2.879 / 0.255           \\
                                                                                                &                                  &                               & $KLD$                                                                    & 1.585 / 0.171           & 1.589 / 0.171           & 2.704 / \textbf{0.244}           \\
                                                                                                &                                  &                               & Ours                                                                    & \textbf{1.319} / \textbf{0.160}           & \textbf{1.388} / \textbf{0.155}           & \textbf{2.440} / 0.226           \\ \hline
\multirow{4}{*}{\begin{tabular}[c]{@{}c@{}}\textit{CNN-based} \\ \textit{Student model}\\ ResNet34 \cite{resnet34}\end{tabular}}  & \multirow{4}{*}{6.64}           & \multirow{4}{*}{4.55}        & -                                                                                       & 0.862 / 0.089           & 1.035 / 0.112           & 1.827 / 0.176           \\
                                                                                                &                                  &                               & $COS$                                                                & 0.829 / 0.088           & 0.943 / 0.107           & 1.694 / 0.164           \\
                                                                                                &                                  &                               & $KLD$                                                                    & 0.771 / \textbf{0.086}           & 0.939 / 0.103           & 1.728 / 0.166           \\
                                                                                                &                                  &                               & Ours                                                                    & \textbf{0.766} / 0.101           & \textbf{0.850} / \textbf{0.096}           & \textbf{1.615} / \textbf{0.161}           \\ \hline
\multirow{4}{*}{\begin{tabular}[c]{@{}c@{}}\textit{D-TDNN-based} \\ \textit{Student model}\\ CAM++\cite{campplus}\end{tabular}}  & \multirow{4}{*}{7.18}           & \multirow{4}{*}{1.72}        & -                                                                                       & 0.718 / \text{\quad-\quad}       & 0.879 / \text{\quad-\quad}       & 1.735 / \text{\quad-\quad}       \\
                                                                                                &                                  &                               & $COS$                                                                & 0.713 / 0.118           & 0.901 / 0.108           & 1.768 / 0.182           \\
                                                                                                &                                  &                               & $KLD$                                                                    & 0.633 / \textbf{0.101}           & 0.790 / 0.093           & 1.572 / 0.159           \\
                                                                                                &                                  &                               & Ours                                                                    & \textbf{0.590} / 0.118           & \textbf{0.735} / \textbf{0.085}           & \textbf{1.494} / \textbf{0.148}           \\ \hline
\end{tabular}

\end{table*}

\section{Experiments Setup}
\subsection{Dataset}
We utilized the VoxCeleb2 dev dataset \cite{vox} for training and evaluated the performance on three test trials, Vox1-O, Vox1-E, and Vox1-H. During training, we applied data augmentation using the MUSAN noise corpus \cite{musan} and RIRs reverberation \cite{rirs}, with a probability of 0.6. 

\subsection{Model}
The teacher model is the SSL-based ASV system \cite{wavlm} combining WavLM Large and ECAPA-TDNN \cite{ecapa}. On the other hand, we utilized various network architectures for our student models including TDNN-based x-vector \cite{xvector}, CNN-based ResNet-34 \cite{resnet34}, and D-TDNN-based CAM++ \cite{campplus}. 

\subsection{Training and Evaluation}
During the training, each audio sample was randomly cropped to a 2-second segment, then 80-dimensional Fbank features were extracted using a frame length of 25 ms and a frameshift of 10 ms. For the classification loss function, we employed the AAM-softmax \cite{aam_softmax} with a scale of 32 and a margin scheduler. In the proposed KD method, from the ablation study in Section \ref{subsec:importance}, the value $\gamma$ in (\ref{eq:final_dkd}) is set to 2.0 in all the remaining experiments. For evaluation, speaker embeddings were scored using cosine similarity and score normalization. Performance is reported on two metrics: Equal Error Rate (EER) and the minimum of the normalized detection cost function (MinDCF) with $P_{target} = 0.01$ and $C_{fa} = C_{miss} = 1$. All experiments are conducted using \textit{Wespeaker} toolkit \cite{wespeaker}. 

\section{Results and Analysis}
\label{sec:result}

\subsection{Results of the proposed method}
Table \ref{tab:main_result} presents a comparison of the performance of teacher and student models trained solely with classification loss, along with the results of different knowledge distillation methods. Although both embedding-level and conventional label-level KD methods outperform student networks trained solely with classification loss, the improvement remains limited. Under the limited number of parameters and floating-point operations (FLOPs), the smallest student model x-vector using DKD emphasizing non-target speakers exhibits the largest improvement of 28.12\% in Vox1-O EER, compared to its baseline trained with classification loss only. Moreover, our proposed method enables the state-of-the-art CAM++ model to further boost its performance with an EER of 0.590\%, while the model's size and FLOPs are respectively 97.73\% and 93.39\% smaller than the teacher model. Lastly, all three student networks trained using our proposed method have a better result than the embedding-level and conventional label-level KD methods, especially in challenging sets like Vox1-E and Vox1-H. This indicates that DKD emphasizing non-target speaker probabilities effectively improves the performance of student models.

\subsection{Ablation Study: The impact of $\mathcal{L}_\text{NSKD}$}
\label{subsec:importance}
We conducted an ablation study on the hyperparameters $\gamma$ in the DKD formula to show how the robustness of the proposed method varies. To save computational cost, we solely present the results of the student model x-vector, which are summarized in Table \ref{tab:ablation}. When emphasizing the $\mathcal{L}_\text{NSKD}$ with non-zero values of $\gamma$, all the results exceed the performance of $\gamma=1-p_\tau^{\mathcal{T}}$, which is equivalent to the result of the conventional label-level knowledge distillation method. It is observed that removing $NSKD$ by assigning $\gamma=0$ obtains a worse result than the conventional label-level knowledge distillation. In alignment with the findings from Section \ref{sec:study}, an increasing value of $\gamma$ leads to better performance, implying the increased significance of $\mathcal{L}_\text{NSKD}$. Notably, the best hyperparameter configuration of $\gamma = 2$ achieved an average of 13\% improvements in EER compared to the conventional label-level knowledge distillation.

\begin{table}[H]
\centering
\caption{Results of x-vector using different $\gamma$ values in Eq.(~\ref{eq:final_dkd}) }
\label{tab:ablation}
\begin{tabular}{l|ccc}
\hline
\multirow{2}{*}{\textbf{$\gamma$}} & \multicolumn{3}{c}{\textbf{EER (\%) / minDCF}}                                                                                                         \\
                                   & \textbf{Vox1-O}  & \textbf{Vox1-E}   & \textbf{Vox1-H}   \\ \hline
$1-p_\tau^{\mathcal{T}}$           & 1.585 / 0.171    & 1.589 / 0.171     & 2.704 / 0.244     \\ \hline
$0.0$                              & 1.622 / 0.152                           & 1.646 / 0.175                                                     & 2.786 / 0.252                           \\
$1.0$                              & 1.463 / 0.166                           & 1.452 / 0.155                                                     & 2.520 / \textbf{0.225} \\
$2.0$                              & \textbf{1.319} / 0.160 & \textbf{1.388} / \textbf{0.155} & \textbf{2.440} / 0.226 \\
$4.0$                              & 1.361 / \textbf{0.143} & 1.415 / 0.156                                                     & 2.511 / 0.229                           \\ \hline
\end{tabular}
\end{table}

\section{Conclusion}
\label{sec:conclusion}
This paper has shown the benefit of leveraging non-target speakers for training automatic speaker verification models. Based on this finding, we modified the conventional label-level KD to emphasize the classification probabilities of non-target speakers, which involves splitting and amplifying the non-target speaker's probabilities during the knowledge distillation process. Experimental results on the VoxCeleb test sets show an average of 13.67\% improvement in EER of the proposed method compared to other knowledge distillation methods across three different architecture student models. 
\section{Acknowledgement}
\label{sec:ack}
The computational work for this article was partially performed on resources of the National Supercomputing Centre, Singapore (https://www.nscc.sg).

\bibliographystyle{IEEEbib}
\bibliography{Template}

\begin{thebibliography}{10}

\bibitem{ssl_app_asr}
Mirco Ravanelli, Jianyuan Zhong, Santiago Pascual, Pawel Swietojanski, Joao
  Monteiro, Jan Trmal, and Yoshua Bengio,
\newblock ``Multi-task self-supervised learning for robust speech
  recognition,''
\newblock in {\em IEEE ICASSP}, 2020, pp. 6989--6993.

\bibitem{ssl_app_kw}
Wei-Tsung Kao, Yuan-Kuei Wu, Chia-Ping Chen, Zhi-Sheng Chen, Yu-Pao Tsai, and
  Hung-Yi Lee,
\newblock ``On the efficiency of integrating self-supervised learning and
  meta-learning for user-defined few-shot keyword spotting,''
\newblock in {\em SLT}, 2023, pp. 414--421.

\bibitem{ssl_app_age_2}
Tarun Gupta, Duc-Tuan Truong, Tran~The Anh, and Chng~Eng Siong,
\newblock ``Estimation of speaker age and height from speech signal using
  bi-encoder transformer mixture model,''
\newblock in {\em Proc. INTERSPEECH}, 2022, pp. 1978--1982.

\bibitem{ssl_app_sv}
Ruijie Tao, Kong Aik~Lee, Rohan Kumar~Das, Ville Hautamäki, and Haizhou Li,
\newblock ``Self-supervised speaker recognition with loss-gated learning,''
\newblock in {\em IEEE ICASSP}, 2022, pp. 6142--6146.

\bibitem{ssl_app_sv2}
Tianchi Liu, Kong~Aik Lee, Qiongqiong Wang, and Haizhou Li,
\newblock ``Disentangling voice and content with self-supervision for speaker
  recognition,''
\newblock in {\em Thirty-seventh Conference on Neural Information Processing
  Systems}, 2023.

\bibitem{emb_kl_kd_1}
Shuai Wang, Yexin Yang, Tianzhe Wang, Yanmin Qian, and Kai Yu,
\newblock ``Knowledge distillation for small foot-print deep speaker
  embedding,''
\newblock in {\em IEEE ICASSP}, 2019, pp. 6021--6025.

\bibitem{emb_kd_2}
Zhiyuan Peng, Xuanji He, Ke~Ding, Tan Lee, and Guanglu Wan,
\newblock ``Label-free knowledge distillation with contrastive loss for
  light-weight speaker recognition,''
\newblock in {\em ISCSLP}, 2022, pp. 324--328.

\bibitem{emb_kd_4}
Xuechen Liu, Md~Sahidullah, and Tomi Kinnunen,
\newblock ``Distilling multi-level x-vector knowledge for small-footprint
  speaker verification,''
\newblock in {\em arXiv preprint arXiv:2303.01125}, 2023.

\bibitem{emb_kd_5}
Jungwoo Heo, Chan yeong Lim, Ju~ho~Kim, Hyun seo Shin, and Ha-Jin Yu,
\newblock ``{One-Step Knowledge Distillation and Fine-Tuning in Using Large
  Pre-Trained Self-Supervised Learning Models for Speaker Verification},''
\newblock in {\em Proc. INTERSPEECH}, 2023, pp. 5271--5275.

\bibitem{kl_kd_2}
Leying Zhang, Zhengyang Chen, and Yanmin Qian,
\newblock ``{Knowledge distillation from multi-modality to single-modality for
  person verification},''
\newblock in {\em Proc. INTERSPEECH}, 2021, pp. 1897--1901.

\bibitem{nontarget_speaker}
Nik Vaessen and David {van Leeuwen},
\newblock ``{Training speaker recognition systems with limited data},''
\newblock in {\em Proc. Interspeech 2022}, 2022, pp. 4760--4764.

\bibitem{nontarget_face}
Yaobin Zhang and Weihong Deng,
\newblock ``Class-balanced training for deep face recognition,''
\newblock in {\em IEEE/CVF CVPRW}, 2020, pp. 3594--3603.

\bibitem{dkd}
Borui Zhao, Quan Cui, Renjie Song, Yiyu Qiu, and Jiajun Liang,
\newblock ``Decoupled knowledge distillation,''
\newblock in {\em IEEE/CVF CVPR}, 2022, pp. 11943--11952.

\bibitem{wavlm}
Sanyuan Chen, Chengyi Wang, Zhengyang Chen, Yu~Wu, Shujie Liu, Zhuo Chen, Jinyu
  Li, Naoyuki Kanda, Takuya Yoshioka, Xiong Xiao, et~al.,
\newblock ``Wavlm: Large-scale self-supervised pre-training for full stack
  speech processing,''
\newblock {\em IEEE Journal of Selected Topics in Signal Processing}, vol. 16,
  no. 6, pp. 1505--1518, 2022.

\bibitem{xvector}
David Snyder, Daniel Garcia-Romero, Gregory Sell, Daniel Povey, and Sanjeev
  Khudanpur,
\newblock ``X-vectors: Robust dnn embeddings for speaker recognition,''
\newblock in {\em IEEE ICASSP}, 2018, pp. 5329--5333.

\bibitem{resnet34}
Hossein Zeinali, Shuai Wang, Anna Silnova, Pavel Mat{\v{e}}jka, and
  Old{\v{r}}ich Plchot,
\newblock ``{BUT} system description to {VoxCeleb} speaker recognition
  challenge 2019,''
\newblock in {\em arXiv preprint arXiv:1910.12592}, 2019.

\bibitem{campplus}
Hui Wang, Siqi Zheng, Yafeng Chen, Luyao Cheng, and Qian Chen,
\newblock ``{CAM++: A Fast and Efficient Network for Speaker Verification Using
  Context-Aware Masking},''
\newblock in {\em Proc. INTERSPEECH}, 2023, pp. 5301--5305.

\bibitem{vox}
Arsha Nagrani, Joon~Son Chung, Weidi Xie, and Andrew Zisserman,
\newblock ``{VoxCeleb}: Large-scale speaker verification in the wild,''
\newblock {\em Computer Speech \& Language}, vol. 60, pp. 101027, 2020.

\bibitem{musan}
David Snyder, Guoguo Chen, and Daniel Povey,
\newblock ``{MUSAN}: {A} {M}usic, {S}peech, and {N}oise {C}orpus,'' 2015,
\newblock arXiv:1510.08484v1.

\bibitem{rirs}
Tom Ko, Vijayaditya Peddinti, Daniel Povey, Michael~L. Seltzer, and Sanjeev
  Khudanpur,
\newblock ``A study on data augmentation of reverberant speech for robust
  speech recognition,''
\newblock in {\em IEEE ICASSP}, 2017, pp. 5220--5224.

\bibitem{ecapa}
Brecht Desplanques, Jenthe Thienpondt, and Kris Demuynck,
\newblock ``{ECAPA}-{TDNN}: Emphasized channel attention, propagation and
  aggregation in {TDNN} based speaker verification,''
\newblock in {\em Proc. INTERSPEECH}, 2020, pp. 3830--3834.

\bibitem{aam_softmax}
Jiankang~Deng et~al,
\newblock ``{ArcFace}: Additive angular margin loss for deep face
  recognition,''
\newblock {\em {IEEE} Transactions on Pattern Analysis and Machine
  Intelligence}, vol. 44, no. 10, pp. 5962--5979, oct 2022.

\bibitem{wespeaker}
Hongji Wang, Chengdong Liang, Shuai Wang, Zhengyang Chen, Binbin Zhang,
  Xu~Xiang, Yanlei Deng, and Yanmin Qian,
\newblock ``Wespeaker: A research and production oriented speaker embedding
  learning toolkit,''
\newblock in {\em IEEE ICASSP}, 2023, pp. 1--5.

\end{thebibliography}

\end{document}